\documentclass[journal]{IEEEtran}
\usepackage[nobreak]{cite}
\usepackage{graphicx}
\usepackage{float}
\usepackage{amsmath}
\usepackage{amssymb}
\usepackage{amsfonts}
\usepackage[printonlyused]{acronym}
\usepackage[capitalise]{cleveref}
\crefformat{equation}{#2(#1)#3}
\Crefformat{equation}{#2(#1)#3}
\usepackage{multirow}
\providecommand{\expe}[1]{\ensuremath{\mathrm{e}^{#1}}}

\def\j{\mathrm{j}}
\def\Haar{\mathop{\mathcal{H}}\nolimits}

\usepackage{url}
\usepackage{tikz}
\usetikzlibrary{calc}

\newcommand{\positiontextbox}[4][]{%
	\begin{tikzpicture}[remember picture,overlay]
		\node[inner sep=3pt, fill=yellow,align=left,draw,line width=1pt,#1] at ($(current page.north west) + (#2,-#3)$) {\parbox{.80\paperwidth}{#4}};
	\end{tikzpicture}%
}

\begin{document}

\title{Time-Modulated Arrays with Haar Wavelets}	
\author{Roberto Maneiro-Catoira,~\IEEEmembership{Member,~IEEE,}
	Julio Br\'egains,~\IEEEmembership{Senior~Member,~IEEE,}\\
	Jos\'e A. Garc\'ia-Naya,~\IEEEmembership{Member,~IEEE,}
	and~Luis Castedo,~\IEEEmembership{Senior~Member,~IEEE}
	\thanks{%
		$^*$ Corresponding author: Jos\'e A. Garc\'ia-Naya (jagarcia@udc.es).
		
		This work has been funded by the Xunta de Galicia (ED431G2019/01), the Agencia Estatal de Investigación of Spain (TEC2016-75067-C4-1-R, RED2018-102668-T) and ERDF funds of the EU (AEI/FEDER, UE).
		
		The authors are with CITIC Research Center \& Department of Computer Engineering,  Universidade da Coru\~na (Univ. of A Coru\~na), Spain. E-mail: roberto.maneiro@udc.es, julio.bregains@udc.es, jagarcia@udc.es, luis@udc.es
		
		Digital Object Identifier 10.1109/LAWP.2020.2986832}
	\thanks{}}

\markboth{IEEE ANTENNAS AND WIRELESS PROPAGATION LETTERS}%
{}
\maketitle

\acrodef{ADC}[ADC]{analog-to-digital converter}
\acrodef{AWGN}[AWGN]{additive white Gaussian noise}
\acrodef{ASK}[ASK]{amplitude-shift keying}
\acrodef{BER}[BER]{bit error ratio}
\acrodef{BF}[BF]{beamforming}
\acrodef{BS}[BS]{beamsteering}
\acrodef{BFN}[BFN]{beamforming network}
\acrodef{DAC}[DAC]{digital-to-analog converter}
\acrodef{DC}[DC]{direct current}
\acrodef{DOA}[DOA]{direction of arrival}
\acrodef{DSB}[DSB]{double sideband}
\acrodef{DFT}[DFT]{discrete Fourier Transform}
\acrodef{ETMA}[ETMA]{enhanced time-modulated array}
\acrodef{FSK}[FSK]{frequency-shift keying}
\acrodef{FT}[FT]{Fourier Transform}
\acrodef{HDWT}[HDWT]{Haar Discrete Wavelet Transform}
\acrodef{HT}[HT]{Hilbert Transform}
\acrodef{ISI}[ISI]{inter-symbol interference}
\acrodef{LMD}[LMD]{linearly modulated digital}
\acrodef{LNA}[LNA]{low noise amplifier}
\acrodef{MBPA}[MBPA]{Multibeam phased-array antenna}
\acrodef{MMIC}[MMIC]{monolithic microwave integrated circuit}
\acrodef{MRC}[MRC]{maximum ratio combining}
\acrodef{MSLL}[MSLL]{maximum side-lobe level}
\acrodef{NMLW}[NMLW]{normalized main-lobe width}
\acrodef{NPD}[NPD]{normalized power density}
\acrodef{PCB}[PCB]{printed circuit board}
\acrodef{PS}[PS]{phase shifter}
\acrodef{PSK}[PSK]{phase-shift keying}
\acrodef{QAM}[QAM]{quadrature amplitude modulation}
\acrodef{RF}[RF]{radio frequency}
\acrodef{RFC}[RFC]{Rayleigh fading channel}
\acrodef{RPDC}[RPDC]{reconfigurable power/divider combiner}
\acrodef{SA}[SA]{simulated annealing}
\acrodef{SER}[SER]{symbol error rate}
\acrodef{SLL}[SLL]{sideband-lobe level}
\acrodef{SNR}[SNR]{signal-to-noise ratio}
\acrodef{SPDT}[SPDT]{single-pole double-throw}
\acrodef{SPMT}[SPMT]{single-pole multiple-throw}
\acrodef{SPST}[SPST]{single-pole single-throw}
\acrodef{SR}[SR]{sideband radiation}
\acrodef{SSB}[SSB]{single sideband}
\acrodef{SSB-TM-MBPA}[SSB TM-MBPA]{single sideband time-modulated multibeam phased-array antenna}
\acrodef{STA}[STA]{static array}
\acrodef{SWC}[SWC]{sum of weighted cosines}
\acrodef{TM}[TM]{time-modulated}
\acrodef{TMA}[TMA]{time-modulated array}
\acrodef{TM-MBPA}[TM-MBPA]{time-modulated multibeam phased-array antenna}
\acrodef{UWB}[UWB]{ultra-wide band}
\acrodef{VGA}[VGA]{variable-gain amplifier}
\acrodef{VPS}[VPS]{variable phase shifter}
\acrodef{VA}[VA]{variable attenuator}

\begin{abstract}
Time-modulated arrays (TMAs) can effectively perform beamsteering over the first positive harmonic pattern by applying progressively delayed versions of stair-step approximations of a sine waveform to the antenna excitations. In this letter, we consider synthesizing such stair-step sine approximations by means of Haar wavelets. Haar functions constitute a complete orthonormal set of rectangular waveforms which have the ability to represent a given function with a high degree of accuracy using few constituent terms. Hence, when they are applied to TMA synthesis, employing single-pole double-throw switches, such a feature leads to an excellent rejection level of the undesired harmonics as well as a bandwidth greater than that supported by conventional TMAs with on-off switches.
\end{abstract}



%
\IEEEpeerreviewmaketitle
	
	\positiontextbox{11cm}{27cm}{\footnotesize \textcopyright 2020 IEEE. This version of the article has been accepted for publication, after peer review. Personal use of this material is permitted. Permission from IEEE must be obtained for all other uses, in any current or future media, including reprinting/republishing this material for advertising or promotional purposes, creating new collective works, for resale or redistribution to servers or lists, or reuse of any copyrighted component of this work in other works. Published version:
	\url{https://doi.org/10.1109/LAWP.2020.2986832}}

\section{Introduction}
\IEEEPARstart{T}{ime-modulated} arrays (\acs{TMA})\acused{TMA} have the ability to perform \ac{BS} by adjusting the on-off instants of the switches that constitute their feeding network. They can be considered as a cost-effective alternative for smart antennas solutions that does not require \acp{VPS}. \ac{TMA} designs, however, have some handicaps such as the control of the unexploited harmonics \cite{Maneiro2017_a,Tong2010}, the presence of mirrored frequency diagrams \cite{Poli2011,Amin_Yao2015}, the transmitted  (and  received)  signal  energy  wasted  during  the zero-state of the switches \cite{Zhu2012,Yang2004b}, and the allowable signal bandwidth due to the spectral overlapping with the signal replicas \cite{Yang2019,Bogdan2019}. When designing \acp{TMA} for \ac{BS} purposes, the aforementioned drawbacks can be alleviated by means of: 
\begin{enumerate}
\item The use of stair-step approximations of time-delayed sine functions --with fundamental frequency $\omega_0$-- as the \ac{TMA} modulating waveforms. This significantly decreases the level of the unexploited \ac{SR}. Stair-step approximations also avoid the energy-absorbing zero-state of the conventional on-off switches and are easily implementable with \ac{SPDT} switches \cite{Maneiro2019a,ManeiroCatoira2019_EUSIPCO}.

\item The use of \ac{SSB} (or complex) \ac{TMA} architectures capable of suppressing power-consuming frequency-mirrored harmonic patterns \cite{Amin_Yao2015, Maneiro2019}, even using amplitude-phase weighting with multiple branches \cite{Yang2019}.
\end{enumerate}

\begin{figure}[ht]
	\centering
	\includegraphics[width=6.5cm]{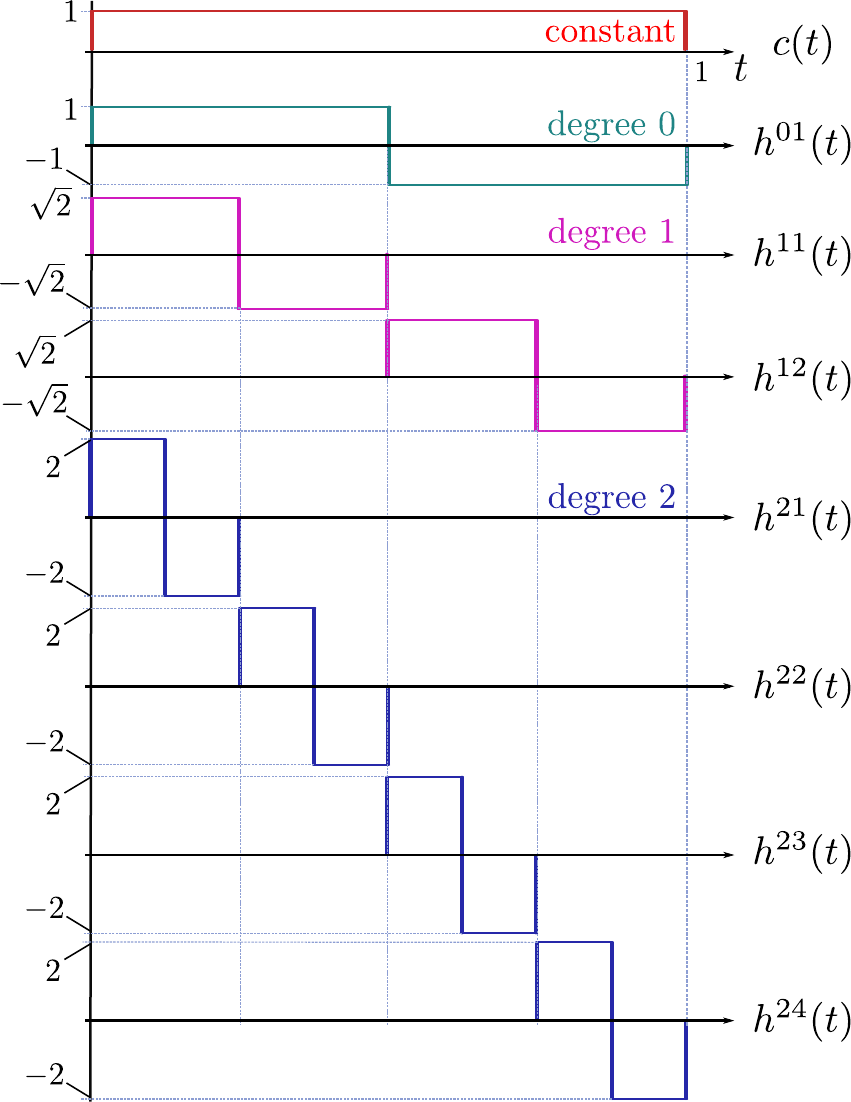}
	\caption{The first eight Haar wavelets $h^{lm}(t)$ with degree $l=\{0,1,2\}$ and order $m=1,\dots, 2^l$.} 
	\label{fig:Haar functions}
	\vspace*{-0.5cm}
\end{figure}

In this letter we will follow a different approach. The approximated time-delayed sine waveforms which modulate the individual \ac{TMA} elements will be synthesized by means of a complete set of orthogonal functions: the Haar wavelets \cite{McLaughlin1969,Shore1973}. As shown in \cref{Sec:Haar functions}, a Haar wavelet $h^{lm}(t)$ is characterized by its degree $l$ and order $m$. Haar wavelets with the same degree $l>0$ are successive time-multiplexed (non overlapped) versions of the corresponding $m=1$ first-order wavelet (see \cref{fig:Haar functions}). Hence, Haar wavelets are well suited for \ac{TMA} synthesis because those with the same degree can be easily generated employing \ac{SPDT} switches, thus enabling a significant complexity reduction.

Accordingly, a given function can be expressed as a linear combination of Haar wavelets whose coefficients are obtained solving integrals similar to those of the Fourier series coefficients, but using the corresponding Haar wavelet instead of sines or cosines. Furthermore, analogously to the \ac{DFT}, the \ac{HDWT} can be used to efficiently compute the Haar coefficients. Indeed, since Haar wavelets may be written in matrix notation by a Haar matrix, when a vector with samples of the waveform to be approximated is given, the calculation of the Haar coefficients is performed by just a matrix product.

The main contribution of this letter is the application of Haar wavelets to the design of \ac{TMA} modulating waveforms to perform  beamsteering  over  the  first  positive  harmonic  pattern.
\section{Haar Wavelets}\label{Sec:Haar functions}
Except for the special case $c(t)=1$ for $0\leq t \leq 1$ (a constant unitary function), Haar wavelets are defined as follows
\begin{align}\label{eq:Haar math definition}
h^{lm}(t)=\begin{cases}
\sqrt{2^l} & \frac{m-1}{2^l}\leq t \leq \frac{m-\frac{1}{2}}{2^l}\\
-\sqrt{2^l} & \frac{m-\frac{1}{2}}{2^l}< t \leq \frac{m}{2^l}\\
0& \text{otherwise}
\end{cases},
\end{align} 
being $l=0,1,2,\dots$ and $m=1,\dots, 2^l$. The degree $l$ denotes a subset having the same number of zero crossings in a given width, ${1}/{2^l}$, and the order $m$ gives the position of the function within this subset. All the members of a subset with the same degree are obtained by shifting the first member along the axis by an amount proportional to its order (see \cref{fig:Haar functions}).

A given continuous function, $f(t)$, within the interval $0\leq t \leq 1$ and repeated periodically outside this interval, can be synthesized from a Haar series as follows \cite{McLaughlin1969}:
\begin{equation}\label{eq:Haar expansion theory}
f(t)=W^0+\sum_{l=0}^{\infty}\sum_{m=1}^{2^l}W^{lm}h^{lm}(t) = W^0+\sum_{l=0}^{\infty}\Haar^l(t)
\end{equation}
and, by virtue of the orthonormality between Haar wavelets, we have that the corresponding Haar wavelet coefficients are
\begin{align}\label{eq:Haar coefficients}
W^0=\int_{0}^{1}f(t)dt \text{ and } W^{lm}=\int_{0}^{1}f(t)h^{lm}(t)dt,
\end{align}
satisfying the extremal of the squared error integral condition
\begin{equation}\label{eq:error}
\lim_{\Gamma\to\infty}\int_{0}^{1} \left|f(t)-\left(W^0+\sum_{l=0}^{\Gamma}\Haar^l(t)\right)\right|^2dt=0.
\end{equation}
If the series expansion in \cref{eq:Haar expansion theory} is truncated at $l=\Gamma$, a finite set of Haar wavelets is considered for the synthesis and a stair-step approximation of $f(t)$ is obtained. Furthermore, the \ac{HDWT} can be interpreted as the mathematical operator that converts a finite sequence of equally-spaced samples of $f(t)$ into a sequence (with the same length) of Haar wavelet coefficients. For numerical handling, we consider a discrete series of $M=2^p$ terms (with $p \in \mathbb{N}$) obtained by sampling $f(t)$ at $M$ equally spaced points $x_k$ over $[0,1)$, with $k \in \Psi=\{1, 2,\dots, M\}$. Hence, the integrals in \cref{eq:Haar coefficients} can be replaced by the finite sums:
\begin{align}\label{eq:Discrete Haar coefficients}
& W^0=\frac{1}{M}\sum_{k=1}^{M}f(x_k) \text{ and } W^{lm}=\frac{1}{M}\sum_{k=1}^{M}f(x_k)h^{lm}(x_k),
\end{align}
with $l \in \Lambda =\{0,\dots, p-1\}$ and $m \in \Xi =\{1,\dots, 2^l\}$. The resulting values for $\{W^0,W^{lm}\}$ constitute the \ac{HDWT} of $f(t)$. Notice that the \ac{HDWT} can be recursively described by a real-valued square matrix considering the Kronecker product (denoted as $\otimes$)
as follows:
\begin{equation}\label{eq:Haar Matrix}
H_M=\frac{1}{\sqrt{2}} \begin{bmatrix}
H_{M/2}& \otimes & \begin{pmatrix}+1 & +1\end{pmatrix} \\
I_{M/2}& \otimes & \begin{pmatrix}+1 & -1\end{pmatrix}
\end{bmatrix},
\end{equation} 
being $(+1 +1)$ and $(+1 -1)$ row vectors, and $I_{M/2}$ the identity matrix of order $M/2$. The iteration starts with $H_1=[1]$, and we easily realize that the first $M$ Haar wavelets (see \cref{fig:Haar functions}) --or rather, samples of such wavelets-- are the rows of $H_M$.
Hence, by considering $f(t)$ in 
$[0,T_0)$, with normalized period $T_0=1$, we can arrange $M=2^p$ equally spaced samples of $f(t)$ in a column vector $\bar{f}_M=[f(x_1),\dots, f(x_M)]^T$ and represent --by virtue of \cref{eq:Discrete Haar coefficients}-- the corresponding \ac{HDWT} of $f(t)$ through the following matrix equation:
\begin{equation}\label{eq:matrix DWT}
\overline{W}_M{=}[W^0 W^{01} W^{11} W^{12} \cdots W^{(p-1)2^{p-1}}]{=}\frac{1}{M}H_M\cdot\overline{f}_M,
\end{equation}
being $\overline{W}_M$ a column vector with the Haar-wavelet coefficients in \cref{eq:Discrete Haar coefficients}, and $H_M$ the \ac{HDWT} matrix in \cref{eq:Haar Matrix} with order $M$.

\begin{figure}[t]
	\centering
	\includegraphics[width=\columnwidth]{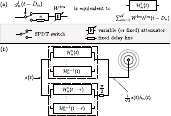}
	\caption{(a) \ac{SPDT} switching architecture capable of generating the term $\sum_{m=1}^{2^l}W^{lm}h^{lm}(t)$ in \cref{eq:u_n(t) segunda} for a given Haar wavelet of degree $l$. Notice that $g_n^l(t-D_n)$ is a square wave with frequency $2^l f_0$ subject to a time delay $D_n$, whereas $\Haar_n^l(t) = \sum_{m=1}^{2^l}W^{lm}h^{lm}(t-D_n)$. The Haar coefficients $W^{lm}$ are implemented with either a variable or a fixed (depending on $l$) attenuator. (b) Generalized architecture for the $n$-th \ac{TMA} element feeding network which synthesizes a Haar wavelet. Note that for the entire array, $N$ time-modulators like this one are needed (plus a $1{:}N$ splitter).} 
	\label{fig:architecture}
\end{figure} 

\section{Time-varying array factor controlled by Haar wavelets}
We propose to apply Haar synthesis to design \acp{TMA} with \ac{BS} capabilities. The idea is to approximate the functions employed for time-modulating the \ac{TMA} excitations (sine waveforms) by means of linear combinations of Haar wavelets (easily implemented with \ac{SPDT} switches and \acp{VA}). Hence, the time-varying array factor is expressed as a function of the Haar coefficients of such functions.

\cref{fig:architecture} shows the proposed feeding architecture for the $n$-th element of a linear \ac{TMA}  with $N$ isotropic elements ($n \in \Theta = \{0,\dots, N-1\}$) requiring only \ac{SPDT} switches, \acp{VA}, and fixed delay lines. In such a feeding network, the excitation of the $n$-th antenna element is time-modulated by the periodic ($T_0$) pulse $h_n(t)=h(t-D_n)$, being $h(t)=f(t)+ \j f(t-\tau)$, where $f(t)$ is an approximation of a sine waveform with fundamental frequency $\omega_0=2\pi f_0=2\pi/T_0$, $\j$ is the imaginary unit, and $D_n$ and $\tau$ are adaptive and fixed (defined beforehand) time delays, respectively.

\begin{figure*}[t]
	\centering
	\includegraphics[scale=0.95]{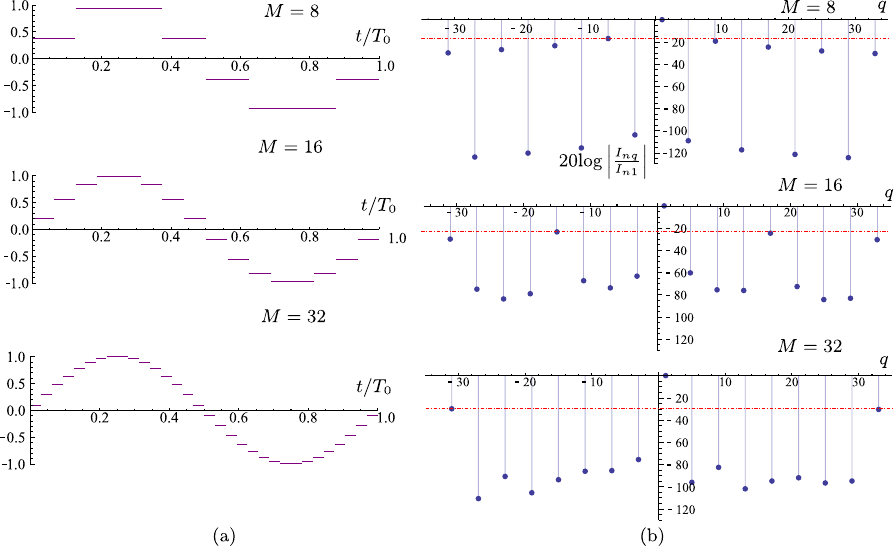}
	\caption{(a) Stair-step approximations of $\sin(2\pi t)$ considering the \ac{HDWT} with $M=8$, $16$, and $32$ equally spaced points. (b) Relative power level of the dynamic excitations of the unexploited harmonics. \cref{tab:Parameters simulations 2} shows the values of the peak \ac{SR} and the maximum signal bandwidth $B_{\text{max}}$.} 
	\label{fig:Stair-step and harmonic plots}
\end{figure*}

\begin{table}[t]
	\caption{Time-varying Haar BFNs for each value of $M$ when a sine waveform is synthesized.}
	\label{tab:Parameters simulations}
	\vspace*{-0.2cm}
	\begin{center}
		\setlength\tabcolsep{0.2em}
		\def\arraystretch{1.3}
		\begin{tabular}{|c|c|c|c|c|}
			\hline
			&{$\Haar^0(t)$ }&{$\Haar^2(t)$}&{$\Haar^3(t)$}&{$\Haar^4(t)$}\\
			\hline \hline
			$M=8$&yes&yes&no&no\\ 
			\hline
			$M=16$&yes&yes&yes&no\\ 
			\hline
			$M=32$&yes&yes&yes&yes\\
			\hline
			Type of attenuator&none&fixed&VA&VA\\ 
			\hline
			Discrete levels (dB)& 0 &-13.6&-20.0,-27.7&-28.5,-29.9,-33.4,-42.5\\ 
			\hline
		\end{tabular}
	\end{center}
	\vspace*{-0.5cm}
\end{table}

\begin{table}[t]
	\caption{Performance comparison between SSB switched TMAs.}
	\label{tab:Parameters simulations 2}
	\vspace*{-0.2cm}
	\begin{center}
		\setlength\tabcolsep{0.2em}
		\def\arraystretch{1.3}
		\begin{tabular}{|c|c|c|c|c|c|}
			\hline
			Reference & Peak SR (dB) & $B_{\text{max}}$ (Hz) & $\eta_\text{TMA}$(\%) &$\eta_\text{mod}$(\%) & $\eta$(\%) \\
			\hline \hline
			\cite{Yang2019}&-25.00&8$f_0$&97.90&20.70&20.27\\ \hline
			\cite{Amin_Yao2015}&-13.98&4$f_0$&91.19&33.33&30.40\\ \hline
			\cite{QChen2019}&-16.90&8$f_0$&94.96&50.00&47.48\\ \hline
			\cite{Maneiro2019a}&-16.90&8$f_0$&96.00&58.00&55.68\\ \hline
			\cite{ManeiroCatoira2019_EUSIPCO}&-23.50&16$f_0$&98.72&50.00&49.36\\ \hline
			Proposed ($M$=32)&-29.80&32$f_0$&99.68&50.00&49.84\\ \hline
		\end{tabular}
	\end{center}
	\vspace*{-0.5cm}
\end{table}

In the synthesis of $f(t)$ in \cref{eq:Haar expansion theory}, we realize that $W^0=0$ because $f(t)$ is an approximation of a pure sine (without direct-current component) and $h^{lm}(t)$ is a periodic ($T_0$) Haar wavelet whose Fourier series expansion is given by
$h^{lm}(t)=\sum_{q=-\infty}^{\infty}G_q^{lm}\expe{\j q\omega_0 t}$,
being $G_{q}^{lm}$ the corresponding Fourier coefficients. By substituting the previous equation into \cref{eq:Haar expansion theory} we have
\begin{align}\label{eq:u_n(t) segunda}
f(t)&=\sum_{l=0}^{p-1}\sum_{m=1}^{2^l}W^{lm}h^{lm}(t)\notag\\
&=\sum_{q=-\infty}^{\infty}\left[ \sum_{l=0}^{p-1}\sum_{m=1}^{2^l}W^{lm}G_{q}^{lm}\right]\expe{\j q\omega_0 t}.
\end{align}
If we select a delay $\tau$ verifying $\omega_0\tau=\pi/2$, then $\expe{-\j q\omega_0\tau}=(-\j )^q$ and, applying the time-shifting property to the Fourier coefficients in \cref{eq:u_n(t) segunda}, we write $h(t)=f(t)+ \j f(t-\tau)$ as
\begin{align}\label{eq:u_n(t-tau}
h(t)=\sum_{q=-\infty}^{\infty}[1-(-\j)^{q+1}]\left[ \sum_{l=0}^{p-1}\sum_{m=1}^{2^l}W^{lm}G_{q}^{lm}\right]\expe{\j q\omega_0 t},
\end{align}
and by applying again the time-shifting property to the Fourier coefficients, we express $h_n(t)=h(t-D_n)$ as
\begin{align}\label{eq:h_n(t)}
h_n(t){=}\sum_{q=-\infty}^{\infty}[1-({-}\j)^{q+1}]\left[ \sum_{l=0}^{p-1}\sum_{m=1}^{2^l}W^{lm}G_{q}^{lm}\right]\expe{-\j q\omega_0D_n}\expe{\j q\omega_0 t}.
\end{align}
Therefore, the  \ac{TMA} element architecture shown in \cref{fig:architecture} leads to the following time-varying array factor (with the term $\expe{\j \omega_c t}$ explicitly included) as a function of the coefficients of the modulating Haar wavelets:
\begin{align}\label{eq:array factor time-domain}
& F(\theta,t) = \frac{\expe{\j q\omega_c t}}{\sqrt{2}}\sum_{n=0}^{N-1}h_n(t)
\expe{\j kz_n\cos\theta}\notag\\
&=\sum_{q=-\infty}^{\infty}\sum_{n=0}^{N-1}\left[ \sum_{l=0}^{p-1}\sum_{m=1}^{2^l}\frac{1-(-\j)^{q+1}}{\sqrt{2}}W^{lm}G_{q}^{lm}\cdot\right.\notag\\
&\cdot\expe{-\j q\omega_0D_n}\expe{\j kz_n\cos\theta}\Bigg]
\expe{\j (\omega_c+q\omega_0) t}=\sum_{q=-\infty}^{\infty}F_q(\theta)\expe{\j (\omega_c+q\omega_0)t},
\end{align} 
\begin{figure*}[!ht]
	\centering
	\includegraphics[scale=1.1]{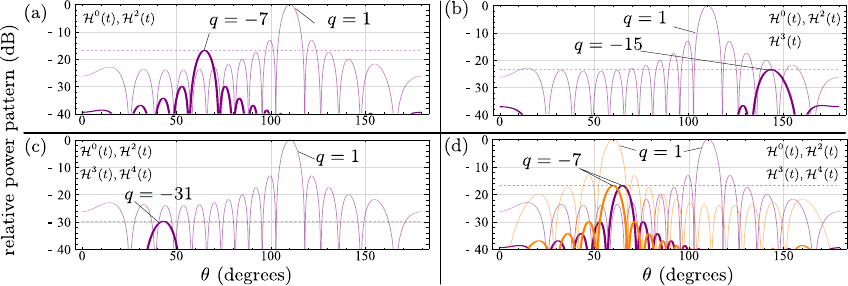}
	\caption{(a), (b), and (c): Power radiated patterns of single-beam \acp{TMA} designed with Haar wavelets for $M=8$, $16$, and $32$ (Haar feeding networks employed are explicitly indicated). Notice that the phase excitations are considered to be progressive, i.e., the useful harmonic beam points to the direction $\theta_0$ (in the example, $\theta_0=110^{\circ}$) when $\omega_0 D_n=\pi n\cos(\theta_0)$, $n \in \Theta$ (see \cref{eq:dynamic excitations}). (d) An advantage of the technique in terms of flexibility is that the single beam architecture employed for $M=32$ can be exploited to generate two independent beams with the features corresponding to $M=8$.} 
	\label{fig:patterns}
\end{figure*}
\noindent where $z_n$ represents the $n$-th array element position on the $z$ axis, $\theta$ is the angle with respect to such a main axis, and $k=2\pi/\lambda_c$ represents the wavenumber for a wavelength $\lambda_c = 2\pi \mathrm{c} /\omega_c$, with $\omega_c$ being the carrier frequency and $\mathrm{c}$ the speed of light. 
Notice that $F_q(\theta)=\sum_{n=0}^{N-1}I_{nq}\expe{\j k z_n\cos\theta}$
is the spatial array factor at the frequency $\omega_c+q\omega_0$ and 
\begin{equation}\label{eq:dynamic excitations}
I_{nq}=\sum_{l=0}^{p-1}\sum_{m=1}^{2^l}\frac{1-(-\j)^{q+1}}{\sqrt{2}}W^{lm}G_{q}^{lm}\expe{-\j q\omega_0D_n}, \ n \in \Theta,
\end{equation}
the corresponding dynamic excitations that synthesize the radiated pattern at such a frequency. As in \cite{Maneiro2019}, note that
\begin{align}\label{eq:valores  (1 + (-j)^{q+1})...}
1-(-\j)^{q+1}=\{2\ q\in \Upsilon; 0\ \text{otherwise}\},
\end{align} 
with $\Upsilon=\{q=4k-3; k\in \mathbb{Z}\}=\{\dots, -7, -3, 1, 5, 9, \dots\}$ and the frequency-mirrored unwanted harmonics are removed (\ac{SSB} feature).
Hence, for a given harmonic pattern of order $q$, the corresponding dynamic excitations $I_{nq}$ in \cref{eq:dynamic excitations} have identical modulus, but they can be endowed with progressive phases for $n \in \Theta$ by selecting $D_n$ with the aim of performing \ac{BS}. The different $D_n$ are easily implemented by means of the switch-on time of the individual Haar wavelets.

\section{Numerical simulations}

We consider a \ac{TMA} with $N=16$ elements spaced $\lambda_c/2$ and the discrete Haar wavelet synthesis of \cref{eq:matrix DWT} applied to $f(t)=\sin(2\pi t/T_0)$ for the cases of $M=8$, $16$, and $32$ equally spaced points in the interval $(0,2\pi]$. As indicated above, we assume a normalized period $T_0=1$. 
\cref{fig:Stair-step and harmonic plots} illustrates the synthesized stair-step approximations of $f(t)$ and the corresponding relative power level of the  dynamic excitations (having identical modulus for $n\in \Theta$) of the unexploited harmonics with respect to the dynamic excitation levels of the useful harmonic at $q=1$.

\cref{tab:Parameters simulations} shows the Haar \acp{BFN} needed for each value of $M$ and the characteristics of the attenuators. $\Haar^0(t)$ does not require attenuation ($W^{lm}=1$ in \cref{fig:architecture}a), $\Haar^2(t)$ employs a fixed attenuator of $-13.6$\,dB, whereas $\Haar^3(t)$ and $\Haar^4(t)$ require \acp{VA} with $2$ and $4$ discrete levels, respectively, which are specified in \cref{tab:Parameters simulations}. \cref{tab:Parameters simulations 2} illustrates the peak \ac{SR}, the maximum signal bandwith ($B_{\text{max}}$), and the efficiencies of the time modulation method \cite{Maneiro2019}: $\eta_{\text{TMA}}$, which accounts for the ability of the \ac{TMA} to filter out and radiate only the useful harmonics; $\eta_{\text{mod}}$, which accounts for the reduction of the total mean power radiated by a uniform static array caused by the insertion of the \ac{TMA} switched feeding network; and $\eta=\eta_{\text{TMA}}\cdot\eta_{\text{mod}}$, which represents the total \ac{TMA} efficiency. The improvement of the peak \ac{SR} and the $B_{\text{max}}$ for $M=16$ (39\% and 100\%, respectively) and $M=32$ (76\% and 300\%, respectively) when compared to that of \cite{QChen2019}, is remarkable. Additionally, we point out that a key difference between \cite{ManeiroCatoira2019_EUSIPCO} and this technique is that Walsh functions occupy an entire period (since they are not time multiplexed like Haar wavelets with the same degree). Hence, each Walsh function must be synthesized by an independent switch, thus increasing the complexity to achieve the same performance.

\cref{fig:patterns}a to \cref{fig:patterns}c show the power radiated patterns and the Haar feeding networks employed for three different values of $M$. These figures evidence the outstanding rejection level of the unwanted harmonics. \cref{fig:patterns}d shows that we can exploit the scheme used for $M=32$ to generate two independent beams. Notice that with feeding networks $\Haar^3(t)$ and $\Haar^4(t)$ (see \cref{fig:architecture}a) we can implement $\Haar^0(t)$ and $\Haar^2(t)$ by modifying the frequency of $g_n^l(t-D_n)$ and the attenuations $W^{lm}$, and governing these two networks by means of square waves with frequencies different than $f_0$ and $4f_0$, respectively. 

It is remarkable that a $b$-bit digital phase shifter, a core element of standard phased arrays, is usually constituted by $2b$ cascaded \ac{SPDT} switches \cite{Analog}, and when $b$ increases, so does both the phase resolution and the insertion losses. With the proposed technique, the phase resolution is independent of the number of switches and, with $2b$ \ac{SPDT} switches, we can implement a Haar \ac{BFN} of order $M=2^{b+1}$. Since $M$ increases with $b$, the total insertion losses of the Haar \ac{BFN} will decrease with $b$ (see $\eta$ values in \cref{tab:Parameters simulations 2}).

\section{Conclusions}
We proposed a novel approach to \ac{TMA} beamsteering based on \ac{HDWT} modulation. Compared to existing switched \acp{TMA}, the method provides better rejection of the undesired harmonics and allows for higher signal bandwidths, although at the expense of a moderate increase of the hardware complexity. 
\ifCLASSOPTIONcaptionsoff
  \newpage
\fi

\vfill



\bibliographystyle{IEEEtran}
%
\bibliography{IEEEabrv,main}

%


\vfill


\end{document}